\documentclass[conference]{IEEEtran}
  \usepackage[left=0.60in, right=0.60in, bottom=0.98in, top=0.63in]{geometry} 
\usepackage{url}
\usepackage[utf8]{inputenc} 
\usepackage[nolist,nohyperlinks]{acronym}

\usepackage[ruled,vlined]{algorithm2e}
\usepackage{amsmath}
\usepackage{amsfonts}
\usepackage{amssymb}
\usepackage{amsthm}
\usepackage{dsfont}
\usepackage{array}
\usepackage{multirow}
\usepackage{lipsum}
\usepackage{mathtools}
\usepackage{cuted}
\usepackage[caption=false]{subfig}
\usepackage{bbm}
\usepackage{siunitx}
\usepackage{bm}
\usepackage{booktabs}

\usepackage{xcolor}

\usepackage{stfloats}
\usepackage{rotating} 
\usepackage{cases}

\usepackage{hyperref}
\usepackage{cleveref}
\usepackage{lineno}

\newcommand{\setS}{\mathcal{S}}

\newcommand{\setT}{\mathcal{T}}

\newcommand{\setF}{\mathcal{F}}

\newcommand{\setE}{\mathcal{E}}
\newcommand{\setX}{\mathcal{X}}
\newcommand{\setY}{\mathcal{Y}}

\newcommand{\setP}{\mathcal{P}}

\DeclareSIUnit \belm {Bm}
\newcommand{\numdBm}[1]{\qty{#1}{\deci\belm}} 

\newcommand{\N}{\ensuremath{\mathbb{N}}}

\newcommand{\R}{\ensuremath{\mathbb{R}}}  

\usepackage{dsfont}

\newcommand{\defeq}{\ensuremath{\triangleq}} 

\Crefname{equation}{Eq.}{Eqs.}
\Crefname{figure}{Fig.}{Figs.}

\usepackage{enumitem}

\newtheorem{definition}{Definition}

\newtheorem{assumption}{Assumption}
\newtheorem{theorem}{Theorem}

\usepackage[draft, authormarkup=none, todonotes={textsize=tiny, textwidth= 40pt}]{changes}
\setdeletedmarkup{} 
\definechangesauthor[name={}, color=blue]{all}
\definechangesauthor[name={}, color=orange]{pz}
\IEEEoverridecommandlockouts
\IEEEpubid{\begin{minipage}{\textwidth}\ \\[50pt]
		{This work has been submitted to the IEEE for possible publication. Copyright may be transferred without notice, after which this version may no longer be accessible.}
\end{minipage}}

\begin{document}
 \bstctlcite{MyBSTcontrol}
\title{
Efficient Integration of~Distributed~Learning Services in Next-Generation Wireless Networks
}

 \author{\IEEEauthorblockN{Paul Zheng\IEEEauthorrefmark{1}, Navid Keshtiarast\IEEEauthorrefmark{2}, Pradyumna Kumar Bishoyi\IEEEauthorrefmark{2},
  Yao Zhu\IEEEauthorrefmark{1},} \IEEEauthorblockN{
 Yulin Hu\IEEEauthorrefmark{3}\IEEEauthorrefmark{1}, Marina~Petrova\IEEEauthorrefmark{2}, and Anke Schmeink\IEEEauthorrefmark{1}}
 \IEEEauthorblockA{\IEEEauthorrefmark{1}Chair of Information Theory and Data Analytics, 
 	RWTH Aachen University, Germany.\\
 Email: $zheng|zhu|schmeink$@inda.rwth-aachen.de
 }
 \IEEEauthorblockA{\IEEEauthorrefmark{2}Mobile Communications and Computing Group, RWTH Aachen University, Germany\\
 Email: $navid.keshtiarast|pradyumna.bishoyi|petrova$@mcc.rwth-aachen.de}
 \IEEEauthorblockA{\IEEEauthorrefmark{3}School of Electronic Information, Wuhan
 	University, China.
 Email: $yulin.hu$@whu.edu.cn}
 \vspace{-.8cm}

 \thanks{
The work of P. Zheng, N. Keshtiarast, P. K. Bishoyi, Y. Zhu, M. Petrova, and A.
Schmeink is supported by the BMBF Germany in the project ``Open6GHub"
under grant 16KISK012. The work of Y. Hu was supported in part by the National Key R\&D Program of China under Grant 2023YFE0206600 and the Fundamental Research Funds for the Central Universities under Grant 2042024kf1006. Y. Hu is the corresponding author.
}
 	  }


\IEEEoverridecommandlockouts

\maketitle

\begin{abstract}
Distributed learning (DL) is considered a cornerstone of intelligence enabler, since it allows for collaborative training without the necessity for local clients to share raw data with other parties, thereby preserving privacy and security. 
Integrating DL into the 6G networks requires coexistence design with existing services such as high-bandwidth (HB) traffic like eMBB. Current designs in the literature mainly focus on communication round (CR)-wise designs that assume a fixed resource allocation during each CR. However, fixed resource allocation within a CR is a highly inefficient and inaccurate representation of the system's realistic behavior. This is due to the heterogeneous nature of the system, where clients inherently need to access the network at different times.  This work zooms into one arbitrary communication round and demonstrates the importance of considering a time-dependent resource-sharing design with HB traffic. We propose a time-dependent optimization problem for minimizing the consumed time and energy by DL within the CR. Due to its intractability, a session-based optimization problem has been proposed assuming a large-scale coherence time. An iterative algorithm has been designed to solve such problems and simulation results confirm the importance of such efficient and accurate integration design.
\end{abstract}
\begin{IEEEkeywords}
Distributed learning, federated learning, 6G networks, eMBB, coexistence design.
\end{IEEEkeywords}

\IEEEpeerreviewmaketitle

\vspace{-.15cm}
\section{Introduction}
\vspace{-.07cm}
Data-driven machine learning techniques have shown substantial potential in tackling intricate problems that traditional optimization struggles with, largely owing to the availability of a huge quantity of data. 
However, the enormous volume of data to be trained on, that is collected or generated by the ever-increasing and diverse range of Internet-of-Things (IoT) devices may contain sensitive and private information about individuals or industries. Sharing this raw information with other parties poses inevitable concerns.

To address privacy and security concerns, a line of distributed learning (DL) frameworks, such as federated learning (FL)~\cite{mcmahan_FL_2017} and split learning~\cite{Gupta_2018_split_learning}, has been proposed. In such frameworks, clients are not required to transmit their data to any third party; instead, only model parameter or intermediate activation output is communicated.
Such DL has drawn immense attention in the past few years and is being considered for integration (or already is integrated) into real-world mobile/edge applications~\cite{Li2020_reviewFL_applications}, e.g., Google Gboard~\cite{hard_2019_FLmobilekeyboard}. 
Due to the limitation of edge devices, the communication quality, training latency, and available energy for training are among the key bottlenecks of such frameworks.
Significant work has been proposed to alleviate this issue~\cite{yang_FLenergy_2021, Xu_2021_LongTermFL, Nguyen_2022_LatencyOpti_blockchainFL, Wu_2024_CostEfficientFLMultiCell, Poposka_2024_DelayMinFLWPT}.
The authors in~\cite{yang_FLenergy_2021} laid the foundation of joint learning accuracy, energy, and time-aware system design for federated edge learning. A long-term design of FL considering that the updates in the later stages of learning are more important is presented in~\cite{Xu_2021_LongTermFL}. Multi-cell latency-and energy-aware FL system designs are proposed in~\cite{Nguyen_2022_LatencyOpti_blockchainFL, Wu_2024_CostEfficientFLMultiCell}. A wireless powered networks-enabled FL has been examined in~\cite{Poposka_2024_DelayMinFLWPT}.
The above work demonstrates the trade-off between learning accuracy, latency and energy, highlighting the importance of an accurate characterization of their relationships.

However, all the above work (and most existing work) considered fixed dedicated wireless resources for DL.  Given the DL process, 
there may exist some idle communication time or periods of lower communication requirements (e.g., for downlink communication), as well as highly bandwidth-demanding periods. Depending on the local computational task and speeds, clients may finish the training and therefore need to transmit their model back to the server at very different times. Giving fixed and dedicated resources to each client of the DL service throughout one round is therefore unreasonable and inefficient as also mentioned in~\cite{Vu_2022_sessionBasedMIMOFL_designs}. 
Since DL is expected to be integrated into the future-generation networks as already considered by 3GPP~\cite{3GPP_AIML_service}, its coexistence with other services such as high-bandwidth (HB) traffic (e.g., eMBB) and URLLC should be studied. 
It is key to remember that unlike other services, where the traffic QoS requirement is typically real-time demanding, thus not ``controllable", DL service traffic is actually real-time tunable, provided that the end-latency and total consumed energy remain within budget, which offers significant room for optimization.

There exists little literature that studies DL coexistence with other services~\cite{Ganjalizadeh_DeviceSelectionURLLCDistributedLearning_2022,FL-nonFL_Farooq_2024}. In~\cite{Ganjalizadeh_DeviceSelectionURLLCDistributedLearning_2022}, the authors study the integration of DL and URLLC services in an industrial network. A risk sensitivity-based formulation for device selection, aimed at minimizing DL training delay while maintaining the required URLLC QoS, is proposed. Further, in~\cite{FL-nonFL_Farooq_2024}, the authors address the co-existence of FL traffic with HB traffic services, considering both half-duplex and full-duplex schemes in a massive MIMO network.
However, no studies have focused on coexistence in the time-dependent resource allocation domain between HB traffic and DL traffic.

In this work, we investigate time-dependent resource allocation for the coexistence of HB traffic services and DL within one communication round (CR). 
For the sake of modeling clarity, this work will use FL as an example of DL. Other types of DL may be designed similarly using similar principles. The aim of the work is to demonstrate the hidden complexity of the resource allocation problem when resources are assumed to be not rigidly fixed, along with the potential large inefficiency gap and completion time estimation error that can occur. It is crucial to exploit the communication idle time and periods of low communication requirements for a highly efficient system. The key contributions of this work are as follows:
\begin{itemize}[leftmargin=11pt]
	 \item We propose a time-dependent resource allocation and computational speed optimization problem to minimize the latency and the consumed energy of CR by vanilla FL while considering certain HB traffic requirements.
\item Due to the potential large model size to be communicated, like~\cite{Vu_2022_sessionBasedMIMOFL_designs}, based on the assumption of large-scale coherence time, we propose an equivalent session-based optimization framework
 that jointly controls the downlink and uplink duration of each session, allocation of RBs of the coexisting DL and HB traffic, and the computational capacity within each session under realistic constraints.
	\item We have proposed an iterative algorithm to solve the non-convex optimization problem. The convexity of the iterative subproblem has been established.
	\item The proposed session-based optimization problem depends on the ordering of the starting time of the uplink sessions. We propose a reasonable heuristic method to obtain an ordering that has been confirmed effective by simulations.
	\item The simulation results confirm the efficiency of the time-dependent design. In resource-constrained systems, the proposed method offers significant improvements in both time and energy objectives compared to the time ``rigid" allocation. It is also compared with considering DL as an HB traffic service, further confirming the necessity of considering DL as a novel service.
\end{itemize}

\vspace{-.15cm}
\section{System Model and Initial Problem Formulation}
\vspace{-.05cm}

Consider a 5G NR system with TTI slots of lengths~$\Delta$~(s).
We consider an existing single-cell system with HB traffic UEs~$\setE$. The DL service is expected to be integrated, we consider FL user equipment (UE) set~$\setF$. In total~$K$ subchannels are available for both services, with a subcarrier spacing of~$B$~(Hz). 

\vspace{-.15cm}
\subsection{High Rate UEs (e.g., eMBB)}
\vspace{-.05cm}
The high rate or HB UEs are performing downlink transmission with~$K_e^{(t)}$ RBs at certain time step~$t$. Under the frequency-flat assumption (when~$K$ is small), the rate can be written as:
\vspace{-.08cm}
\begin{equation}
(\forall e\in\setE)\quad 	r^{(t)}_{e} = K_e^{(t)}B\log_2\Big(1+\gamma_{e}^{(t)} \Big) ,
 \vspace{-.1cm}
\end{equation}
where~$\gamma_{e}^{(t)}= \frac{P^{(d)}h_{e}^{(t)2}}{BN_0}$ is the SNR of UE~$e\in\setE$; $h_e^{(t)}$ the channel coefficient of UE~$e$ and $N_0$ the AWGN noise level. The BS is assumed to have a constant downlink power~$P^{(d)}$ at each RB.

A fair rate allocation for HB traffic UEs needs to be ensured. The requirement is for all HB traffic UEs to have the time average rate above a threshold~$\theta$~(bit/s):
\begin{equation}
\min_{e\in\setE}\frac{1}{T}\sum_{t=1}^{T} r_e^{(t)} \geq \theta,
\label{eq: time_raw_eMBB}
\end{equation}
where~$T>0$ is the ending time of the considered FL CR. 

\subsection{Vanilla Federated Learning}
\vspace{-.05cm}
To highlight the potential of time-dependent resource allocation, this work will focus on vanilla FL so that learning accuracy is not a subject of concern here. This ensures the generality of the work since it means that it can also be applied in a more advanced designed FL framework.

Vanilla FL, i.e. FedAvg~\cite{mcmahan_FL_2017}, consists of repeating iteratively the following steps:
\begin{enumerate}[leftmargin=16pt]
    \item The BS selects~$S$ UEs~$\setS\subset\setF$ to participate in the FL training of the next CR, and broadcast the current global model to those UEs.
    \item After receiving the global model, each UE~$s\in\setS$ trains for $I_s$ epochs with its own local dataset.
    \item Each UE sends back the locally trained model once the local training is done.
    \item After receiving all model updates from UEs, the average of the model updates is computed at the BS and is considered as the global model for the next CR.
\end{enumerate}
In this work, we focus on resource allocation within each CR. We assume therefore an arbitrary client selection~$\setS\subset\setF$ and the step~$4)$ is ignored since it is not impacted by wireless resource allocations.



\noindent\textbf{Downlink Session:} The downlink communication of FL is done via broadcasting. 
The downlink rate of FL UE~$s\in\setS$ at time step~$t$ writes similarly as downlink HB traffic:
\vspace{-.03cm}
\begin{equation}
r^{(t)}_{s,dl} = K_{dl}^{(t)}B\log_2\Big(1+\gamma_{s,dl}^{(t)} \Big),
\end{equation}
where~$K_{dl}^{(t)}$ the number of RBs given to the downlink broadcasting; 
$\gamma_{s,dl}^{(t)} = \frac{P^{(d)}h_s^{(t)2}}{N_0}$ the downlink SNR of UE~$s$; $h_s^{(t)}$ the channel coefficient of UE~$s$ at time step~$t$. 

With a model size of~$D$ bits, the downlink session will last for~$\tau_{s,dl}$ for UE~$s\in\setS$:
\vspace{-.30cm}
\begin{equation}
 \quad\quad\quad\quad\quad   \sum_{t=1}^{\tau_{s,dl}} r_{s,dl}^{(t)} = D.
    \label{eq: time_raw_dlcompletion}
\end{equation}

\noindent\textbf{Local Training Update:} Each UE asynchronously starts the local training on its own after receiving the whole model ($D$ bits). The local training lasts for a time denoted by~$\tau_{s}^{(cp)}$ that is tunable by the computational capacity~$f_{s}\in(0, f_{s,\max}]$~\cite{yang_FLenergy_2021}:
\begin{equation}
	\tau^{(\text{cp})}_{s} = \frac{I_sC_s\Theta_s}{f_{s}},
\end{equation}
where~$C_s$ (cycles/sample) the number of CPU cycles required for training one sample data at UE~$s$; $I_s$ the number of local epochs and $\Theta_s$ is the local data size where the number of iterations (given a fixed epoch) depends on..

The energy consumed on the local computation of UE~$s$ can be written as:
\begin{equation}
	E^{(\text{cp})}_{s}(f_{s}) = \kappa I_s C_s \Theta_sf_{s}^2,
\end{equation}
with~$\kappa>0$ is the effective switched capacitance~\cite{yang_FLenergy_2021}.

\noindent\textbf{Model uplink update:} After the local training, each UE again on its own, asynchronously requests to transmit the updated model to BS for averaging. 


In downlink, BS serves each RB with the same amount of power given sufficient power of transmission of BS. However, the transmission power of edge devices is limited and the more RB is given to a UE,
the less SNR it has on each subchannel:
\vspace{-.1cm}
\begin{equation}
r_{s,ul}^{(t)} = K_{s,ul}^{(t)} B \log_2\Bigg(1+\frac{p_{s,ul}^{(t)}h_{s}^{(t)} }{K_{s,ul}^{(t)}B N_0}\Bigg),
\label{eq: uplink rate (time raw)}
\vspace{-.06cm}
\end{equation}
where~$K_{s,ul}^{(t)}$ the number of RBs used by UE~$s$ for the uplink transmission at TTI~$t$;  $p_{s,ul}^{(t)}$ the transmission power. Each UE~$s$ is subject to a maximum transmit power $P_{\max}$, i.e., $p_{s,ul}^{(t)}\in[0, P_{\max}]$.

The uplink transmission time~$\tau_{s,ul}$ writes as:
\begin{equation}
(\forall s\in\setS)\quad \sum_{t=\tau_{s,dl}+\tau_s^{(cp)}+1}^{\tau_{s,dl}+\tau_s^{(cp)}+\tau_{s,ul}} r_{s,ul}^{(t)} = D,
    \label{eq: time_raw_uplink_completion}
\end{equation}
where reasonably, the allowed starting time of the uplink transmission is when the downlink and the local computation have been completed: $\tau_{s,dl}+\tau_s^{(cp)}$.
The resulting consumed energy in the uplink communication consists:
\begin{equation}
(\forall s\in\setS)\quad E_s^{(cm)} =\Delta \sum_{t=\tau_{s,dl}+\tau_s^{(cp)}+1}^{\tau_{s,dl}} p_{s,ul}^{(t)}.
\end{equation}

The overall latency of the CR is characterized by the slowest UEs:
\begin{equation}
T = \max_{s\in\setS}\{ \tau_{s,dl} + \tau^{(cp)}_{s}(f_{s}) + \tau_{s,ul}\}.
\end{equation}

The overall consumed energy of the CR for UE~$s$ is denoted as $
E_s^{(tot)} = E_s^{(cp)}+E_s^{(cm)}.$


\vspace{-.15cm}
\subsection{General Problem Formulation}
\vspace{-.05cm}
We aim at efficiently allocating limited RBs between HB traffic and FL traffic over time, the uplink transmission power, and the device computational capacity to minimize the total latency of one FL CR with a certain energy penalty while the HB traffic requirement is satisfied.  The problem can be formulated as follows:
\begin{subequations}
	\label{pb: orig_pb}
	\begin{align}
		\!\!\!\!	\min_{\{K_{e}^{(t)}, K_{s,dl}^{(t)}, K_{s,ul}^{(t)}, p_{s,ul}^{(t)},f_s\}_{\forall e,s,t}, } \!\!\!\!\!\!\!\!\!\!\! & T + \sum_{s}\lambda_s (E_s^{(cp)}+E_s^{(cm)}), \\
		\text{s.t.} \quad\quad\quad& \eqref{eq: time_raw_eMBB}, \eqref{eq: time_raw_dlcompletion},\eqref{eq: time_raw_uplink_completion} \\
		&\hspace{-2.5cm} \forall t,\ \sum_{e\in\setE}K_e^{(t)} + K_{s,dl}^{(t)} + \sum_{s}K_{s,ul}^{(t)} \leq K,
  \label{cons: Res_sharing (t step)}\\[-.22cm] 
		&\hspace{-2.8cm} \forall s,e,t, \ f_s\in (0, f_{s,\max}], \ K_{e}^{(t)}, K_{s,dl}^{(t)}, K_{s,ul}^{(t)}\in\N.
	\end{align}
	\vspace{-.1cm}
\end{subequations}
The constraint~\eqref{cons: Res_sharing (t step)} denotes the resource sharing of each traffic at each time step.
\vspace{-.15cm}
\subsection{Rigid Resource Allocation (as baseline) and ``Static" Channel Assumption}
\label{sec: rigid}
\vspace{-.05cm}
Typically~$D$ as the model parameter size to be transmitted is big and the training latency is also in the order of second, while the channel coherence time is usually in the order of TTI from 1 to 30~ms. It makes sense to design an allocation that accounts for the time average rate. We assume that CR endures within the large-scale channel coherence time, where the time-average channel characteristics remain constant. 

Most current wireless FL designs focus on the long-term overall FL performance (across the whole FL training process over multiple CRs) and therefore consider a rigid RB allocation within each CR. The rigid formulation for solving the problem~\eqref{pb: orig_pb} is as follows with the variables set $\setX_{rig}=\{(K_{dl}\in[0,K'], K_{s,ul}\in\R_+, p_{s,ul}\in [0, P_{\max}], \tau_{s,cp}\in[\tau_{s,\min},+\infty])_s\}$:
\begin{subequations}
	\begin{align}
		\min_{\setX_{rig}} \quad & \max_s \Big\{\frac{D}{r_{s,dl}(K_{dl})}+ \tau_{s,cp} + \frac{D}{r_{s,ul}(K_{s,ul}, p_{s,ul})}\Big\} \notag \\
		& +\sum_s\lambda_s\Big[\kappa \frac{\alpha^3\Theta_{s}^3}{\tau_{s,cp}^2} + \frac{p_{s,ul}D}{r_{s,ul}(K_{s,ul}, p_{s,ul})} \Big]  \\[-.1cm]
		\text{s.t.}\quad&  \sum_s K_{s,ul}\leq K' \\[-.3cm]
		& \forall s, \tau_{s,cp} \geq \max_{s'}\Big\{\frac{D}{r_{s',dl}(K_{dl})}\Big\},
        \label{cons: dl/ul separation (rigid)}
	\end{align}
\end{subequations}
where~$K' = K - K_{HB}$ with~$K_{HB}>0$ the constant RB that needs to be allocated to HB traffic to satisfy the constraint; the uplink and downlink separation is ensured by~\eqref{cons: dl/ul separation (rigid)}.

\vspace{-.15cm}
\subsection{Session-Aware RB allocation}
\vspace{-.05cm}
\subsubsection{Motivation}

\begin{figure}[t]
    \centering
    \subfloat[Heterogeneous\label{fig: illu_hetero}]{
    \includegraphics[width=0.49\linewidth, trim=5 0 0 5]{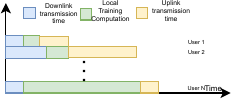}}
    \subfloat[Homogeneous\label{fig: illu_homo}]{
\includegraphics[width=0.49\linewidth, trim=5 0 0 5]{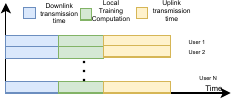}}
    \caption{Example Illustration of the System Time-wise RB allocation for homogeneous and heterogeneous system}
    \label{fig: illustration_heterogeneous}
    \vspace{-.6cm}
\end{figure}
%

The fixed resource allocation across the whole CR can only be accurate if the system is homogeneous, i.e., if the channel strength and the computational capacity and load of all UEs are identical, in this case, a fixed allocation across the whole CR can be an accurate and efficient allocation strategy, e.g., see an illustration in~\Cref{fig: illu_homo}.
When the system is heterogeneous, i.e., when the local dataset size or the computational capacity is very different, the time when the uplink UEs are ready to transmit their local update can vary significantly. As an example in~\Cref{fig: illu_hetero}, at the time when UE~1 is ready to do the uplink transmission, it can actually take much more bandwidth. 
The design is more complex when the energy effect is taken into account. If the network right after UE~1 finishes its transmission is not congested, then maybe the training of UE~1 should be slower to save more energy. If it is congested afterward, then UE~1 should try to finish its load as much as possible right before the congestion starts.

%
%
%
%

\subsubsection{Session-aware RB allocation}
Resource allocation in wireless systems must be adjusted in response to the arrival or end of new traffic demands. In the context of FL, the downlink session only consists of each UE finishing its transmission. The uplink session may consist of a new UE finishing its local training so it is ready for the uplink transmission or the UE finishing its uplink transmission. Here, we only assume the starting time as the boundary of a session in the uplink phase.
\begin{definition}[Session]
The boundary of a \textbf{session} is determined by the moment a UE completes its transmission in the downlink phase and the moment it can initiate its transmission in the uplink phase.
\end{definition}
The RB allocation strategy will remain constant throughout each session.
Due to the potentially large data size to compute and high BS transmit power, the downlink broadcasting is generally much shorter than the local computations. We therefore make the following realistic assumption:
\begin{assumption}
	\label{assumption: separate dl/ul}
Uplink communications start only after the end of all the downlink broadcasting communication.
\end{assumption}

Since the downlink sessions always follow the channel strength ordering, we order the UEs indices in the descending order of channel strength:
\begin{equation}
h_1 \geq h_2\geq \cdots \geq h_S.
\end{equation}
There are $S!$ potential combinations of sessions ordering for uplink sessions since the local computing tasks and the energy requirement can be potentially heterogeneous.

Denoting the downlink and uplink session indices respectively  $\ell', \ell=1,\ldots, S$.
The delay of each CR becomes: $
	T = \sum_{\ell'} t_{\ell'}^{(dl)}+ T_{idle} + \sum_{\ell} t_{
		\ell}^{(ul)},$
where $t_{\ell'}^{(dl)}$ is the duration of downlink session~$\ell'$, i.e., ending by the time for the UE~$\ell'$ to finish its transmission; $t_{\ell}^{(ul)}$ is the duration of uplink session~$\ell$, i.e., for~$\ell=1,\ldots, S-1$, starting by the time for the UE~$\sigma(\ell)$ to be ready for its transmission (finishes its local computation). The last session $\ell=S$ ends after all UEs finish their uplink communication tasks; $T_{idle}$ denotes the idle time between the last UE finishes the downlink transmission and the time the fastest UE finishes its computation and is ready for uplink communication. The HB traffic UEs can take full resources during the idle time. 
This idle time is essential for the universality of the session-based problem formulation.  If we consider a substantial computational task, it is unreasonable to force the last downlink UE to wait for completion at the moment the fastest UE can initiate its uplink transmission. The existing session-based work~\cite{Vu_2022_sessionBasedMIMOFL_designs} lacks consideration of this idle time.

By definition of the downlink session duration, it has to verify that the duration of $s$-th UE downlink communication is the sum of all the sessions until~$s$: $
	\tau_{s}^{(dl)} = \sum_{\ell'\leq s} t_{\ell'}^{(dl)}.$
The following inequality has to hold due to the separation of downlink and uplink phase and the definition of $T_{idle}\geq 0$:
\begin{equation}
	\tau_{S}^{(dl)}+ T_{idle}= \sum_{\ell'} t_{\ell'}^{(dl)}+ T_{idle} =  \tau_{\sigma(1)}^{(dl)} + \tau_{\sigma(1)}^{(cp)}  .
	\label{eq: sigma 1 time constraint, T_idle}
\end{equation}
By definition of duration of uplink sessions, the following satisfies for~$\ell=1,\ldots,S-1$: $
	t_{\ell}^{(ul)}=  \tau_{\sigma(\ell+1)}^{(dl)}+\tau_{\sigma(\ell+1)}^{(cp)} - \tau_{\sigma(\ell)}^{(dl)}-\tau_{\sigma(\ell)}^{(cp)}.$
With telescopic sum, for all $s=1,\ldots,S-1$, the sum of the uplink session until the $s$ sessions is:
\begin{equation}
\begin{aligned}
\hspace{-.19cm}	\sum_{\ell\leq s} t_{\ell}^{(ul)}&= \tau_{\sigma(s+1)}^{(dl)}+\tau_{\sigma(s+1)}^{(cp)} - \tau_{\sigma(1)}^{(dl)}-\tau_{\sigma(1)}^{(cp)}\\[-.35cm]
&=  \sum_{\ell'\leq \sigma(s+1)} t_{\ell'}^{(dl)}+\tau_{\sigma(s+1)}^{(cp)} -\sum_{\ell'} t_{\ell'}^{(dl)}- T_{idle}.
\end{aligned}
\end{equation}
We notice that the equality holds for $s=0$ as well with~\eqref{eq: sigma 1 time constraint, T_idle}.
The computation speed constraint can therefore be fully represented by the session durations, for~$s=0,\ldots, S-1$:
\begin{equation}
\tau_{\sigma(s+1)}^{(cp)} = \sum_{\ell\leq s} t_{\ell}^{(ul)}+\hspace{-.3cm}\sum_{\ell'>\sigma(s+1)}\hspace{-.3cm} t_{\ell'}^{(dl)} + T_{idle}\geq \tau_{\sigma(s+1),\min}^{(cp)},
\label{cons: computation time (session_pb)}
\end{equation}
with $\tau_{s,\min}^{(cp)}$ the fastest computation time calculated with the maximal computational capacity~$f_{s,\max}$.

Denote other session-based variables: $K_{\ell'}^{(dl)}$ the number of RB allocated to the~$\ell'$ the downlink session, $K_{e, \ell'}^{(dl)}$ for HB traffic UE~$e$, the downlink communication of UE~$\ell'$ has to finish at the end of downlink session~$\ell'$:
\begin{equation}
	\sum_{i\leq \ell'} r_{\ell'}^{(dl)}(K_{i}^{(dl)})t_{i}^{(dl)} \geq D.
	\label{cons: dl completion (session_pb)}
\end{equation}
The RB sharing during the downlink phase can be written as:
\begin{equation}
	\forall \ell',\ K_{\ell'}^{(dl)} + \sum_eK_{e,\ell'}^{(dl)}\leq K.
	\label{cons: dl RB (session_pb)}
 \vspace{-.1cm}
\end{equation}

All UEs have to finish their transmission at the end of this CR. UE~$\sigma(s)$ can only start its transmission at the $s$-th uplink session, therefore the following expression holds:
\begin{equation}
	\sum_{\ell\geq s}r^{(ul)}_{\sigma(s),\ell}(K_{\sigma(s),\ell}^{(ul)}, p_{\sigma(s),\ell}^{(ul)})t_{\ell}^{(ul)} \geq D,
	\label{cons: ul completion (session_pb)}
\end{equation}
where $r^{(ul)}_{\sigma(s),\ell}$ is the uplink rate function of UE~$\sigma(s)$ (uplink UE starting time ranking at~$s$th place)  at the uplink session~$\ell$, with~$\ell\geq s$; $K_{\sigma(s),\ell}^{(ul)}$ and $p_{\sigma(s),\ell}^{(ul)}$ the RB and the power allocated to UE~$\sigma(s)$ at the uplink session~$\ell$.

The RBs again are shared with HB traffic UEs, at each session~$\ell$,
\begin{equation}
	\sum_{s\leq \ell}K_{\sigma(s),\ell}^{(ul)} + \sum_eK_{e,\ell}^{(ul)}\leq K,
	\label{cons: ul RB (session_pb)}
\end{equation}
with~$K_{e,\ell}^{(ul)}$ the number of RB allocated to~$e$-th HB traffic UE.
With the introduced session-based variables that replace~$\tau_s^{(cp)}$, the total energy consumed for FL by UE~$\sigma(s)$ with~$s=1,\ldots,S$ can be written as:
\begin{equation}
E_{\sigma(s)}^{(tot)} \!=\! \sum_{\ell\geq s }p_{\sigma(s),\ell}^{(ul)}t_{\ell} \!+\!  \frac{\kappa\alpha^3\Theta_{\sigma(s)}^3}{(\!\!\sum\limits_{\ell\leq s-1} \!\!\!\! t_{\ell}^{(ul)}\!+\! \hspace{-.1cm}\sum\limits_{\ell'>\sigma(s)}\hspace{-.3cm} t_{\ell'}^{(dl)} \!+ T_{idle} )^2}.
\end{equation}

For all~$e$, the average HB traffic rate over the whole process is equal to:
\begin{equation}
\hspace{-.2cm}	\frac{1}{T} \!\sum_{t=1}^Tr_e^{(t)}\!  =\! \frac{\sum_{\ell'} r_{e,\ell'}^{(dl)}t_{\ell'}^{(dl)} \!+\! T_{idle} r_{e,\max} \!+\!\sum_{\ell} r_{e,\ell}^{(ul)}t_{\ell}^{(ul)} }{ \sum_{\ell'} t_{\ell'}^{(dl)} + T_{idle}+ \sum_{\ell} t_{\ell}^{(ul)} }.
\end{equation}

The HB traffic requirement~\eqref{eq: time_raw_eMBB} can be therefore written as:
\begin{multline}
	\forall e\in\setE,\ \sum_{\ell'} r_{e,\ell'}^{(dl)}t_{\ell'}^{(dl)} + T_{idle} r_{e,\max} +\sum_{\ell} r_{e,\ell}^{(ul)}t_{\ell}^{(ul)} \\[-.33cm]
\geq \theta \Big( \sum_{\ell'} t_{\ell'}^{(dl)}+ T_{idle} + \sum_{\ell} t_{\ell}^{(ul)} \Big).
	\label{cons: eMBB (session_pb)}
	\vspace{-.2cm}
\end{multline}
\vspace{-.15cm}
\section{Session-based Problem Formulation}
\vspace{-.05cm}

\subsection{Session-based Problem Formulation}
\vspace{-.05cm}
Combining all the constraints and the reformulation considerations, the problem is as follows:
\begin{subequations}
\label{pb: session-based problem}
	\begin{align}
		\!\min\  & \sum_{\ell'} t_{\ell'}^{(dl)}+ T_{idle}+ \sum_{\ell} t_{\ell}^{(ul)} + \sum_{s}\lambda_{\sigma(s)} E_{\sigma(s)}^{(tot)}
  \label{obj: session-based problem}
  \\[-.1cm]
		\text{s.t.}\ \  & \eqref{cons: eMBB (session_pb)}, \eqref{cons: dl completion (session_pb)}, \eqref{cons: ul completion (session_pb)}, \eqref{cons: dl RB (session_pb)}, \eqref{cons: ul RB (session_pb)}, \eqref{cons: computation time (session_pb)}
      \\[-.1cm]
      & \forall s,\ \forall\ell,\   p_{\sigma(s), \ell}\in[0,P_{\max}] 
      \label{cons: power (session_pb)}
        \\[-.1cm]
		& t_{\ell'}^{(dl)}, t_{\ell}^{(ul)}, T_{idle}, K_{s,\ell}^{(ul)}, K_{\ell'}^{(dl)}, K_{e,\ell'}^{(dl)},K_{e,\ell}^{(ul)}   \geq 0.
  \label{cons: positivity (session_pb)}
	\end{align}
\end{subequations}
Denoting the feasible set of problem variables as:  $\setX = \{(t_{\ell'}^{(dl)}, t_{\ell}^{(ul)}, T_{idle}, K_{s,\ell}^{(ul)}, K_{\ell'}^{(dl)}, K_{e,\ell'}^{(dl)},K_{e,\ell}^{(ul)}, p_{s, \ell})\}$.
The resulting problem has reduced the problem variable dimension by~$S$, since now the computational capacity is represented solely by the relationships of the duration variables.

\vspace{-.1cm}
\subsection{Non-Convexity Handling}
\vspace{-.05cm}
The problem is nonconvex due to the product term in the computing energy in the objective, and in the product between the time and rate in the transmission requirement constraints~\eqref{cons: dl completion (session_pb)},\eqref{cons: ul completion (session_pb)},\eqref{cons: eMBB (session_pb)}. The computing energy is to minimize while the completions constraints are to guarantee a minimum, so the descent would go in the maximizing direction. They need to be handled differently.

\subsubsection{Maximizing Product}


The product consists of either the RB allocation with the durations as in~\eqref{cons: eMBB (session_pb)}, \eqref{cons: dl completion (session_pb)}, or the concave uplink rate expression w.r.t. power and RB allocation (perspective of function $x\mapsto \log(1+x)$) with the duration.
Each product can be seen as the quotient of a concave function and a convex function. 
With the development in fractional programming, we employ the well-known \textit{quadratic transform}~\cite{Shen_FP_quadraticTransform_2018} to handle the product terms in order to obtain a good enough stationary point with an iterative algorithm.

Using quadratic transform for handling product terms, given whatever concave function~$X:x\mapsto X(x)$ to multiply with certain duration variable~$t$, the product term can be transformed as:
\vspace{-.1cm}
\begin{equation}
\hspace{-.17cm}g(x,t)\!\defeq\! X(x)t \!=\! \min_y\! \Big(\!2y\sqrt{X(x)} - \!\frac{y^2}{t}\!\Big)\! \defeq\! \min_y \hat{g}(x,\!t,\!y).
\label{eq: quadratic transform}
\end{equation}
The variable~$y$ is introduced as an auxiliary variable. The transform has the advantages of:
\begin{itemize}[leftmargin=11pt]
    \item Equivalent solutions: $(x^*, t^*)$ maximizes of~$g$ if and only if $(x^*, t^*, y^*)$ maximizes~$ \hat{g}$ for chosen~$y^*$.
    \item Equivalent objective: as already stated in~\eqref{eq: quadratic transform}, for any~$(x,t)$, the equality holds with~$g(x,t)=\hat{g}(x,t, y^*)$ with~$y^*=\arg\min \hat{g}(x,t,y) = \sqrt{X(x)}t$, 
\end{itemize}

Using the transform, we introduce a slack variable for each product term in each non-convex constraint as follows while denoting the resulting constraint~$(x)$ with the notation~$\widehat{(x)}$: 
\begin{itemize}[leftmargin=11pt]
\item $\widehat{\eqref{cons: eMBB (session_pb)}}$:  $y^{(dl)}_{e, \ell'}$ and $y^{(ul)}_{e,\ell}$ in~\eqref{cons: eMBB (session_pb)};
\item $\widehat{\eqref{cons: dl completion (session_pb)}}$:  $y^{(dl)}_{\ell'}$ in~\eqref{cons: dl completion (session_pb)};
\item $\widehat{\eqref{cons: ul completion (session_pb)}}$: $y^{(ul)}_{s,\ell}$ in~\eqref{cons: ul completion (session_pb)}.
\end{itemize}


The slack variables updates will be detailed together in the section~\Cref{sec: algorithm}.

\subsubsection{Minimizing Product Term}
The product term to minimize is the communication energy which is the product of the communication power and duration of communication. We aim to find a tight convex approximation. Using the principle of Majorization-Minimization, a tight convex upper bound can be found, for all~$s,\ell$:
\begin{equation}
\hspace{-.17cm}p_{\sigma(s),\ell}^{(ul)}t_{\ell}^{(ul)}\leq \frac{p_{\sigma(s),\ell}^{(ul)2}}{2y^{(E)}_{s,\ell}} + \frac{y^{(E)}_{s,\ell}t_{\ell}^{(ul)2}}{2} \defeq \hat{h}_{s,\ell}(p_{\sigma(s),\ell}^{(ul)}, t_{\ell}^{(ul)}).
\end{equation}
with~$y_{s,\ell}^{(E)} =\hat{p}_{\sigma(s),\ell}^{(ul)} / \hat{t}_{\ell}^{(ul)} $. The function $\hat{h}_{s,\ell}$ is convex and the inequality is tight at the point $(\hat{p}_{\sigma(s),\ell}^{(ul)},\hat{t}_{\ell}^{(ul)})$. Denote the approximated total energy $\hat{E}^{(tot)}_{\sigma(s)} = E^{(cp)}_{\sigma(s)} + \sum_{\ell}\hat{h}_{s,\ell}$.


\vspace{-.15cm}
\subsection{Algorithm}
\vspace{-.05cm}
\label{sec: algorithm}
Given any feasible point of the problem $X\in\setX$, the updates conducted on the auxiliary variables in set~$\setY=\{(y_{embb,\ell'}^{(dl)}, y_{embb,\ell}^{(ul)}, y^{(dl)}_{\ell'}, y_{s, \ell}^{(ul)}, y_{s,\ell}^{(E)})\}\subset \R_{++}^{3S+S(S+1)}$ are:
\begin{equation}
	\begin{cases}
		\forall \ell',e, \ &y_{e,\ell'}^{(dl)} = \sqrt{K_{embb,\ell'}^{(dl)}}t_{\ell'}^{(dl)},\ y^{(dl)}_{\ell'} = \sqrt{K_{\ell'}^{(dl)}}t_{\ell'}^{(dl)}\\
		\forall \ell,e, \ &y_{e,\ell}^{(ul)} = \sqrt{K_{embb,\ell}^{(ul)}}t_{\ell}^{(ul)}\\
		\forall s,\forall\ell,\ &y_{s, \ell}^{(ul)} = \sqrt{r_{\sigma(s), \ell}^{(ul)}}t_{\ell}^{(ul)},\ y_{s,\ell}^{(E)}=\frac{p_{\sigma(s),\ell}^{(ul)}}{t_{\ell}^{(ul)}}.\\
		\vspace{-.8cm}
	\end{cases}
	\label{eq: update on y}
\end{equation}

The resulting transformed problem~$(\setT\setP)$ after all the transformations given $Y\in\setY$ is:
\begin{subequations}
	\label{pb: session-based problem: convex iterative subproblem}
	\begin{align}
	\hspace{-.2cm}(\setT\setP)\!\!:	\!\min  & \sum_{\ell'} t_{\ell'}^{(dl)}\!\!+\! T_{idle}\!+\! \sum_{\ell} t_{\ell}^{(ul)} \!+\! \sum_{s}\lambda_{\sigma(s)} \hat{E}_{\sigma(s)}^{(tot)}
		\\[-.1cm]
		\text{s.t.}\ \  & \widehat{\eqref{cons: eMBB (session_pb)}}, \widehat{\eqref{cons: dl completion (session_pb)}}, \widehat{\eqref{cons: ul completion (session_pb)}}, 
		\eqref{cons: dl RB (session_pb)}, \eqref{cons: ul RB (session_pb)}, \eqref{cons: computation time (session_pb)},\eqref{cons: power (session_pb)},\\[-.1cm]
		& \hspace{-1.5cm}t_{\ell'}^{(dl)}, t_{\ell}^{(ul)}, T_{idle}, K_{s,\ell}^{(ul)}, K_{\ell'}^{(dl)}, K_{embb,\ell'}^{(dl)},K_{embb,\ell}^{(ul)}   \geq 0.
		\label{cons: positivity (session_pb)}
	\end{align}
\end{subequations}

\begin{theorem}
Given fixed auxiliary variables $Y\in\setY$, the subproblem~$(\setT\setP)$ is a convex optimization problem.
\end{theorem}

The proof is omitted here for brevity; the convexity is mainly due to the reformulation and the fact that the uplink rate expression is the perspective function of the logarithmic function and, therefore, jointly concave.
The proposed algorithm to solve the optimization problem~\eqref{pb: session-based problem} given an arbitrary ordering~$\sigma$ is detailed in~\Cref{algo: iterative giving ordering}. It guarantees to converge to a stationary point.

\vspace{-.1cm}
\begin{algorithm}\small
	\SetAlgoLined
\textbf{Initialize:} $X_0\in\setX$, $T_0 = \infty$. \\
	
\For{$n = 1,\ldots$}{
\underline{\emph{Update of the auxiliary variables~$Y$}}\\
Update $Y_{n}$ according to~\eqref{eq: update on y} based on~$X_{n-1}$\;
\underline{\emph{Update of the original variables~$X$}}\\
Solve the convex optimization problem~$(\setT\setP)$: given~$Y_{n}$, obtain~$X_n$ with achieved optimum~$T_n$ \;
\underline{\emph{Stopping criterion}}\\
\If{$\|T_n-T_{n-1}||/\|T_n\| \leq \varepsilon$ or $n \geq n_{\max}$}{Stop loop}
}
\KwResult{$X_n$ and~$T_n$}
\caption{Iterative algorithm for solving~\eqref{pb: session-based problem}}
\label{algo: iterative giving ordering}

\end{algorithm}

\vspace{-.1cm}
\begin{theorem}[\cite{Shen_FP_quadraticTransform_2018}]
The sequence~$(X_n, Y_n)_{n\in\N}$ of~\Cref{algo: iterative giving ordering} converges to a stationary point of problem~\eqref{pb: session-based problem}.
\end{theorem}


%

\subsection{Heuristic Ordering: Rigid-based Ordering}
\label{sec: heuristic Ordering}
An ordering~$\sigma$ of the uplink session starting time, i.e., end of UE computation, is required for the above-developed algorithm. We choose to use the result of the rigid resource allocation in~\Cref{sec: rigid} as a heuristic for this ordering. The heuristic has been confirmed via simulations in~\Cref{sec: simul_verify}.
\vspace{-.1cm}
\section{Simulations}
\vspace{-.12cm}
\subsection{Simulation Settings}
\vspace{-.05cm}
\subsubsection{Settings}
It is considered that at this CR, $S=10$ FL UEs are participating in the training in a resource-constrained wireless system with~$K=10$ RBs coexisting with~$20$ HB traffic UEs. All UEs, FL and HB traffic UEs are uniformly distributed in the cell of radius of $\SI{50}{\meter}$. 
The ``long-term" static average channel is only subject to free-space path loss. 
The minimum HB traffic rate requirement among all UEs of HB traffic is $\SI[per-mode = symbol]{600}{\kilo\byte\per\second}$. We assume that FL UEs are training a neural network of model parameter size of~$D=\SI{100}{\mega\bit}$ on a dataset of the same size and dimension of Cifar-10~\cite{cifar}, i.e., total of 60000 RGB images of size $32\times 32$ with floating points in $32$-bits, distributed among UEs. The 60000 images are assumed to be distributed among the~$S$ UEs according to uniform random ratios. The computation cycle required for one sample $C_s$ is calculated as $15$ cycles per bit~\cite{Do_2022_DRLUAVFL_FLcomputEnergyRef} multiplied by the number of bits contained in one data sample. In this work, the network-wide energy penalty is considered uniform so assume~$\lambda_s=\lambda=0.05$ for all~$s\in\setS$. The complete system parameters are detailed in the~\Cref{tab: param}.

\begin{table}\scriptsize
	\vspace{-.4cm}
	\centering
	\caption{Parameter values used in simulations}
	\label{tab: param}
	\vspace{-.1cm}
	\begin{tabular}{|c|c||c|c|}
		\hline
		\textbf{Parameter} & \textbf{Value} &\textbf{Parameter} & \textbf{Value}\\
		\hline
		$S$ & 10 & $K$ & 10\\\hline
		$N_0$ & -174 dBm/Hz& $|\setE|$ & 20\\\hline
		$P^{(d)}$& \numdBm{30} & $\theta$ & \SI[per-mode = symbol]{600}{\kilo\byte\per\second}\\\hline
		$P_{\max}$ & \numdBm{23} &  $B$ & \SI{60}{\kilo\hertz}\\\hline
		freq & \SI{3.5}{\giga\hertz} &  $D$ & \SI{100}{\mega\bit}\\\hline
		{$\kappa$} &  {$10^{-28}$}&  {$f_{\max}$} &  {2 GHz}\\ \hline
		{$C_s$} &  15x32x32x3x32 &  {$I_s$} &  {20} \\ \hline 
		$\lambda$ & 0.05 & & \\ \hline
	\end{tabular}
	\vspace{-.4cm}
\end{table}
\subsubsection{Baselines}


The baseline methods to compare are listed as follows.
\begin{itemize}[leftmargin=11pt]
    \item Time-uniform \textit{rigid} RB allocation: detailed in~\Cref{sec: rigid}. It is the key baseline since most existing literature is based on this.
    \item Consider FL as an HB traffic: max-sum-rate (\textit{MSR}) and max-min-rate (\textit{MMR}), to show the importance of having a dedicated service class than HB traffic. Note the exact energy planning in this case is not possible.
\end{itemize}

\vspace{-.15cm}
\subsection{Algorithm Convergence + Effect of ranking}
\begin{figure*}
\begin{minipage}[t]{0.33\textwidth}
	\centering
	\includegraphics[width=\linewidth, trim=0 0 0 0]{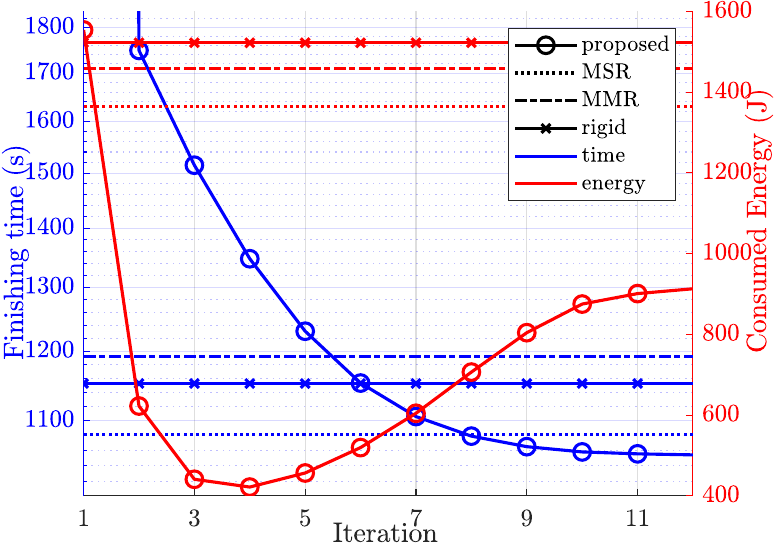}
	\vspace{-.7cm}
	\caption{Algorithm convergence}
	\label{fig: convergence}
\end{minipage}
\begin{minipage}[t]{0.33\textwidth}
	\centering
	\includegraphics[width=0.88\linewidth, trim=0 5 0 -5]{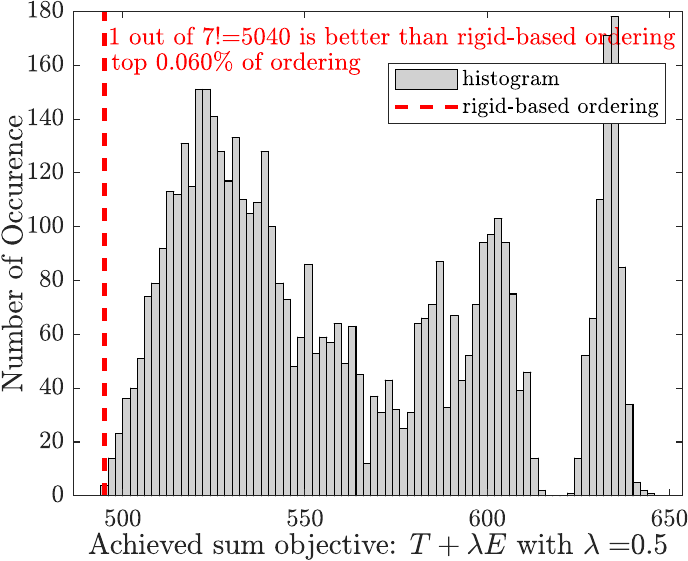}
		\vspace{-.30cm}
	\caption{Rigid ranking evaluation with~$S=7$}
	\label{fig: ranking_confirm}
\end{minipage}
\begin{minipage}[t]{0.33\textwidth}
	\centering
	\includegraphics[width=0.92\linewidth]{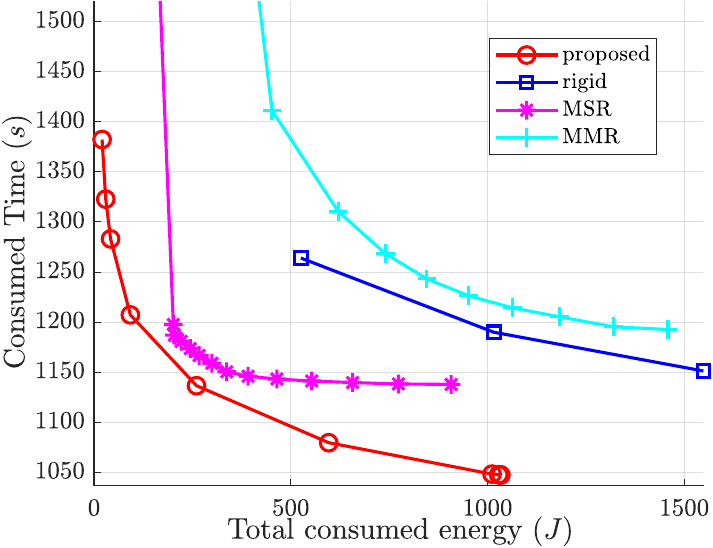}
		\vspace{-.35cm}
	\caption{Energy-time Pareto front}
	\label{fig: Pareto front}
\end{minipage}
\vspace{-.7cm}
\end{figure*}

\vspace{-.05cm}
\label{sec: simul_verify}
The proposed solution consists of an iterative algorithm. Its convergence and performance are verified in~\Cref{fig: convergence}. We observe that the initial point of the proposed algorithm has a finishing time worse than all other methods in the figure (actually out of the figure's scale), but with~8 iterations, it achieves the best-performing method in terms of both criteria and the consumed energy starts to decrease while the latency decreases due to the small penalty value of~$\lambda=0.05$. At the end of iterations, only \textit{MSR }is comparable in terms of finishing time to the proposed method, but the proposed method is only using about \SI{63}{\percent} of energy than \textit{MSR}. All other baselines, i.e., \textit{MMR} and \textit{rigid} have all largely inferior performance in both objectives. We especially observe that \textit{rigid} achieves the worse performance in both objectives, further confirming the necessity of the proposed session-based optimization framework to more accurately estimate the latency and the energy of FL.

As specified in~\Cref{sec: heuristic Ordering}, the proposed method uses the ranking given by the rigid formulation. Its performance is shown in the histogram of~\Cref{fig: ranking_confirm}. It shows that rigid-based ordering does not necessarily imply the best achievable ordering but is among the top ones (here, the second best).

\vspace{-.15cm}
\subsection{Performance Evaluation}
\vspace{-.05cm}
Due to the multi-objective nature of the problem, the energy-time Pareto front is shown in~\Cref{fig: Pareto front}. It can be observed that the proposed method has a large gain compared to \textit{rigid} formulation and is better than \textit{MSR} and \textit{MMR} across all Pareto front. 

\begin{figure}
	\vspace{-.3cm}
	\subfloat[{$K$ influence}\label{fig: Ecstr K}]{\includegraphics[width=0.476\linewidth, trim=0 -15 0 15]{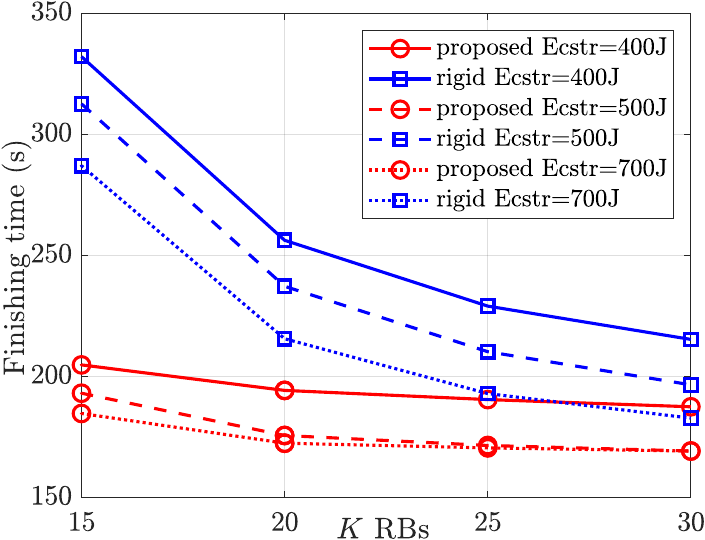}
 		}
	\subfloat[ $\theta$ influence\label{fig: Ecstr theta}]{\includegraphics[width=0.49\linewidth, trim=0 0 0 0]{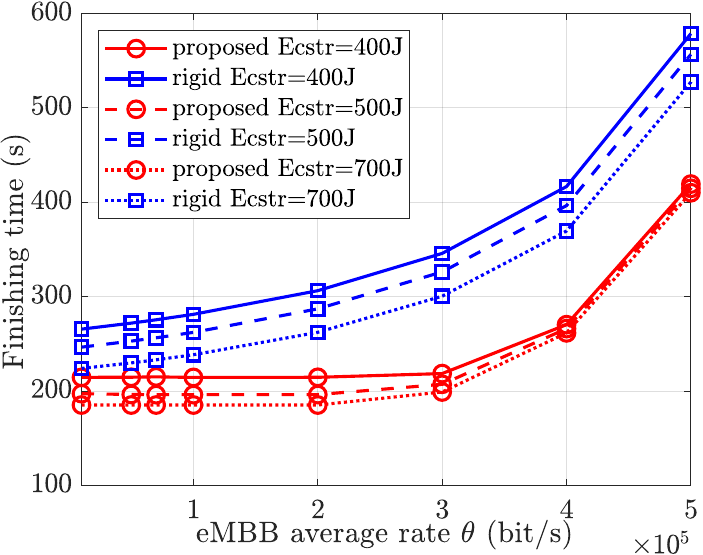}
}
\vspace{-.15cm}
\caption{Parameter influence when fixed energy constraint is imposed.}
\label{fig: params eval}
\vspace{-.55cm}
\end{figure}

The effect w.r.t. the number of RBs~$K$ and~$\theta$ is shown in~\Cref{fig: params eval}. It can be seen that \textit{rigid}-based allocation is more easily changing when the system becomes more constrained, whether for~$K$ becomes smaller, or~$\theta$ larger, or energy constraint becomes smaller.

%






\vspace{-.45cm}
\section{Conclusion}
\vspace{-.05cm}
In this work, we have explored the efficient integration of DL services within next-generation wireless networks, particularly focusing on the coexistence of DL and HB traffic. The proposed time-dependent resource allocation framework, formulated as a session-based problem, addresses the inefficiencies associated with fixed resource assignments in CRs. By considering the heterogeneous nature of local computational tasks and capacities and channel strength, we demonstrated significant improvements in both latency and energy consumption.
Simulation results validate the proposed method. This work underscores the importance of flexible time-dependent resource management strategies for efficiently designing and accurately estimating the consumed latency and energy when enabling DL in future wireless networks. The effect of system heterogeneity on the framework and its impact on the system stragglers will be the subject of future work.



\vspace{-.44cm}
\bibliographystyle{IEEEtran}
\bibliography{related_works.bib}

\begin{thebibliography}{10}
\providecommand{\url}[1]{#1}
\csname url@samestyle\endcsname
\providecommand{\newblock}{\relax}
\providecommand{\bibinfo}[2]{#2}
\providecommand{\BIBentrySTDinterwordspacing}{\spaceskip=0pt\relax}
\providecommand{\BIBentryALTinterwordstretchfactor}{4}
\providecommand{\BIBentryALTinterwordspacing}{\spaceskip=\fontdimen2\font plus
\BIBentryALTinterwordstretchfactor\fontdimen3\font minus
  \fontdimen4\font\relax}
\providecommand{\BIBforeignlanguage}[2]{{%
\expandafter\ifx\csname l@#1\endcsname\relax
\typeout{** WARNING: IEEEtran.bst: No hyphenation pattern has been}%
\typeout{** loaded for the language `#1'. Using the pattern for}%
\typeout{** the default language instead.}%
\else
\language=\csname l@#1\endcsname
\fi
#2}}
\providecommand{\BIBdecl}{\relax}
\BIBdecl

\bibitem{mcmahan_FL_2017}
H.~B. McMahan, E.~Moore, D.~Ramage, S.~Hampson, and B.~A.~Y. Arcas,
  ``Communication-efficient learning of deep networks from decentralized
  data,'' in \emph{AISTATS}, Fort Lauderdale, Florida, USA, 2017.

\bibitem{Gupta_2018_split_learning}
O.~Gupta and R.~Raskar, ``Distributed learning of deep neural network over
  multiple agents,'' \emph{J. Netw. and Computer Applications}, vol. 116, pp.
  1--8, 2018.

\bibitem{Li2020_reviewFL_applications}
L.~Li, Y.~Fan, M.~Tse, and K.-Y. Lin, ``A review of applications in federated
  learning,'' \emph{Computers \& Industrial Engineering}, vol. 149, p. 106854,
  2020.

\bibitem{hard_2019_FLmobilekeyboard}
A.~Hard, K.~Rao, R.~Mathews, S.~Ramaswamy, F.~Beaufays, S.~Augenstein,
  H.~Eichner, C.~Kiddon, and D.~Ramage, ``Federated learning for mobile
  keyboard prediction,'' 2019, arXiv:1811.03604.

\bibitem{yang_FLenergy_2021}
Z.~Yang, M.~Chen, W.~Saad, C.~S. Hong, and M.~Shikh-Bahaei, ``Energy efficient
  federated learning over wireless communication networks,'' \emph{{IEEE}
  Trans. Wireless Commun.}, vol.~20, no.~3, pp. 1935--1949, 2021.

\bibitem{Xu_2021_LongTermFL}
J.~Xu and H.~Wang, ``Client selection and bandwidth allocation in wireless
  federated learning networks: A long-term perspective,'' \emph{{IEEE} Trans.
  Wireless Commun.}, vol.~20, no.~2, pp. 1188--1200, 2021.

\bibitem{Nguyen_2022_LatencyOpti_blockchainFL}
D.~C. Nguyen, S.~Hosseinalipour, D.~J. Love, P.~N. Pathirana, and C.~G.
  Brinton, ``Latency optimization for blockchain-empowered federated learning
  in multi-server edge computing,'' \emph{{IEEE} J. Sel. Areas Commun.},
  vol.~40, no.~12, pp. 3373--3390, 2022.

\bibitem{Wu_2024_CostEfficientFLMultiCell}
T.~Wu, Y.~Qu, C.~Liu, H.~Dai, C.~Dong, and J.~Cao, ``Cost-efficient federated
  learning for edge intelligence in multi-cell networks,'' \emph{{IEEE/ACM}
  Trans. Netw.}, vol.~32, no.~5, pp. 4472--4487, 2024.

\bibitem{Poposka_2024_DelayMinFLWPT}
M.~Poposka, S.~Pejoski, V.~Rakovic, D.~Denkovski, H.~Gjoreski, and
  Z.~Hadzi-Velkov, ``Delay minimization of federated learning over wireless
  powered communication networks,'' \emph{{IEEE} Commun. Lett.}, vol.~28,
  no.~1, pp. 108--112, 2024.

\bibitem{Vu_2022_sessionBasedMIMOFL_designs}
T.~T. Vu, H.~Q. Ngo, M.~N. Dao, D.~T. Ngo, E.~G. Larsson, and T.~Le-Ngoc,
  ``Energy-efficient massive {MIMO} for federated learning: Transmission
  designs and resource allocations,'' \emph{IEEE Open J. Commun. Society},
  vol.~3, pp. 2329--2346, 2022.

\bibitem{3GPP_AIML_service}
``Study on {5G} system support for {AI/ML-based} services,'' 3GPP, Technical
  Specification Group Services and System Aspects, TR 23.700-80, Dec. 2022.

\bibitem{Ganjalizadeh_DeviceSelectionURLLCDistributedLearning_2022}
M.~Ganjalizadeh, H.~S. Ghadikolaei, D.~Gündüz, and M.~Petrova, ``Device
  selection for the coexistence of {URLLC} and distributed learning services,''
  2022, arXiv:2212.11805.

\bibitem{FL-nonFL_Farooq_2024}
M.~Farooq, T.~T. Vu, H.~Q. Ngo, and L.-N. Tran, ``Massive {MIMO} for serving
  federated learning and non-federated learning users,'' \emph{{IEEE} Trans.
  Wireless Commun.}, vol.~23, no.~1, pp. 247--262, 2024.

\bibitem{Shen_FP_quadraticTransform_2018}
K.~Shen and W.~Yu, ``Fractional programming for communication systems—{Part}
  {I}: Power control and beamforming,'' \emph{{IEEE} Trans. Signal Process.},
  vol.~66, no.~10, pp. 2616--2630, 2018.

\bibitem{cifar}
A.~Krizhevsky, ``Learning multiple layers of features from tiny images,'' Tech.
  Rep., 2009.

\bibitem{Do_2022_DRLUAVFL_FLcomputEnergyRef}
Q.~V. Do, Q.-V. Pham, and W.-J. Hwang, ``Deep reinforcement learning for
  energy-efficient federated learning in {UAV}-enabled wireless powered
  networks,'' \emph{IEEE Commun. Letters}, vol.~26, no.~1, pp. 99--103, 2022.

\end{thebibliography}

\end{document}